\documentclass[prd,aps,preprintnumbers,amsmath,amssymb,showpacs,showkeys,nofootinbib]{revtex4}
\usepackage{bm} 
\usepackage{epsfig}

\begin{document}
\title{Dispersion representations and anomalous singularities of the triangle diagram} 
\author{Wolfgang Lucha$^{a}$, Dmitri Melikhov$^{a,b}$ and Silvano Simula$^{c}$}
\affiliation{
$^a$ Institute for High Energy Physics, Austrian Academy of Sciences, Nikolsdorfergasse 18, A-1050, Vienna, Austria\\
$^b$ Nuclear Physics Institute, Moscow State University, 119992, Moscow, Russia\\
$^c$ INFN, Sezione di Roma III, Via della Vasca Navale 84, I-00146, Roma, Italy}
\date{\today}
\begin{abstract}
We discuss dispersion representations for the triangle diagram 
$F(p_1^2,p_2^2,q^2)$, the single dispersion 
representation in $q^2$ and the double dispersion representation in $p_1^2$ and $p_2^2$, with  
special emphasis on the appearance of the anomalous singularities and the anomalous
cuts in these representations. 
For the double dispersion representation in $p_1^2$ and $p_2^2$, 
the appearance of the anomalous cut in the region $q^2>0$ is demonstrated, 
and a new derivation of the anomalous double spectral density is given. 
We point out that the double spectral representation is particularly suitable for applications 
in the region of $p_1^2$ and/or $p_2^2$ above the two-particle thresholds. 
The dispersion representations for the triangle diagram in the nonrelativistic limit 
are studied and compared with the triangle diagram of the nonrelativistic field theory. 
\end{abstract}
\maketitle
\section{Introduction}
The triangle diagrams have many applications in quantum field theory: 
they give the radiative corrections to the form factors of a relativistic 
particle, e.g., quark or electron; they describe the amplitudes of radiative and leptonic decays of
hadrons, e.g., $\pi^0\to\gamma\gamma$; they provide essential contributions to the amplitudes of
hadronic decays, such as $K\to 3\pi$; they give the main contribution to the weak and
electromagnetic form factors of relativistic bound states. Also, these diagrams are responsible 
for one of the most interesting phenomenon of quantum field theory --- for quantum anomalies. 

\begin{figure}[b]
\begin{center}
\epsfig{file=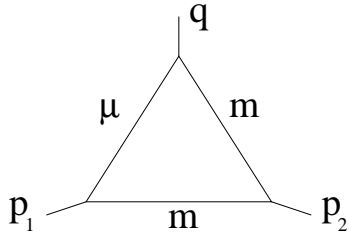,width=4.5cm} 
\caption{\label{fig:0.1} 
The Feynman diagram $F(p_1^2,p_2^2,q^2)$.}
\end{center}
\end{figure}

In this paper we study and compare various spectral representations for the one-loop triangle form-factor Feynman diagram 
with spinless particles in the loop (Fig.~\ref{fig:0.1})\footnote{We
note that the inclusion of spin essentially does not
change the analysis and may be easily done.}
\begin{eqnarray}
\label{f}
F(q^2,p_1^2,p_2^2)=\frac{1}{(2\pi)^4 i}\int  
\frac{dk}{(m^2-k^2-i0)(\mu^2-(p_1-k)^2-i0)(m^2-(p_2-k)^2-i0)}, \qquad q=p_1-p_2.
\end{eqnarray}
The function $F$ is easily calculable in the Euclidean region of all spacelike external momenta 
but has complicated analytic properties in the Minkowski space relevant for
the description of processes with real particles. To handle these processes, dispersion
representations of the diagram are known to be very efficient. 

The application of the dispersion 
representations to the triangle diagram has a long history (see \cite{fronsdal,burton} and references 
therein). An essential feature of the triangle diagram is the appearance of the 
anomalous threshold in a single spectral representation, e.g., in $q^2$ \cite{karplus}: 
the anomalous threshold is located below the normal threshold which is related to the 
possible physical intermediate states in the unitarity relation. 
As a result, mainly the anomalous singularity 
determines the properties of the triangle diagrams in the region of small $q^2$. 
The location of the
anomalous threshold is given by the Landau rules \cite{landau}. 

The double spectral representation in $p_1^2$ and $p_2^2$ for the case of the decay kinematics 
$0<q^2<(\mu-m)^2$ also has an anomalous contribution, which is, however, of a different 
kind than the one in the single representation in $q^2$. Both anomalous contributions 
have a similar origin (they are related to the 
motion of a branch point of the integrand from the unphysical sheet onto the physical sheet through 
the normal cut and the corresponding modification of the integration contour), but the location of the 
anomalous threshold in the double spectral representation is not given by the Landau rules. 
In the decay region $0<q^2<(\mu-m)^2$, the anomalous threshold lies above the normal
threshold, and the anomalous piece dominates the triangle diagram for $q^2\simeq(\mu-m)^2$. 

An exhaustive analysis of the single and the double dispersion representations of the triangle diagram for all
values of the external and the internal masses can be found in \cite{fronsdal}. 

We discuss here the single and the double dispersion representations of the triangle diagram, 
with the emphasis on the properties of the anomalous contributions. We point out that in many 
cases the application of the double spectral 
representation in $p_1^2$ and $p_2^2$ is technically much simpler than the application of the 
single representation in $q^2$.

We start, in Section \ref{sect:ii}, with the case of particles of the same mass in the loop. 
We illustrate the appearance of the anomalous cut in the single spectral representation in $q^2$ 
for $p_1^2>0$, $p_2^2>0$, and $p_1^2+p_2^2\ge 4m^2$. This spectral representation 
has a rather complicated form especially for complex values of $p_1^2$ and $p_2^2$. 
We then discuss the double spectral representation in $p_1^2$ and
$p_2^2$. This representation is very simple for $q^2<0$ and contains only the normal cut. This makes the 
application of the double spectral representation particularly convenient for the analysis of processes 
described by the triangle diagram for timelike $p_1$ and $p_2$ 
in the region $p_1^2+p_2^2\ge 4m^2$ and for higher overthreshold values of $p_1^2$ and $p_2^2$. 

In Section \ref{sect:iii}, we discuss the double spectral representation in 
$p_1^2$ and $p_2^2$ for the case of particles of different masses in the loop.
We give a new derivation of the anomalous contribution to the double spectral representation 
which emerges for the decay kinematics $0<q^2<(\mu-m)^2$. 
This derivation is much simpler than the one known in the literature \cite{braun,melikhov} 
and opens a possibility to consider double spectral representations in the production 
region $q^2>(\mu+m)^2$, which otherwise represents a very complicated technical problem. 
Again, the double spectral representation in $p_1^2$ and $p_2^2$ provides a very convenient 
tool for considering processes at overthreshold values of the variables $p_1^2$ and $p_2^2$, 
relevant for the decay processes, such as, e.g., $K\to 3\pi$ decays. 

In Section \ref{sect:iv}, we consider the double dispersion representation for the triangle diagram 
in the region of the external variables near the thresholds 
\begin{eqnarray}
\label{kin1}
p_1^2=(\mu+m-\epsilon_1)^2,\quad p_2^2=(2m-\epsilon_2)^2,\quad \epsilon_{1,2}\ll m,\mu 
\end{eqnarray}
and for the momentum transfer near the zero recoil 
\begin{eqnarray}
\label{kin2}
q^2=(\mu-m)^2-2m(\mu+m)u^2, \quad u^2\simeq \Lambda/m, \quad \Lambda\ll m. 
\end{eqnarray}
In this region we construct the nonrelativistic expansion of the triangle diagram $F$
and compare it with the triangle diagram of the nonrelativistic field theory $F_{\rm NR}$. 
For the latter we obtain the double dispersion representation in 
$\epsilon_1$ and $\epsilon_2$, the ``binding energies'' of the initial and final states. 
Interestingly, the nonrelativistic expansion of the triangle diagram $F$ is quite different 
for the case of equal masses in the loop and for the decay case $\mu>m$:  
In the case of equal masses and for $q^2<0$, the anomalous cut is absent in the double dispersion representation, 
and the nonrelativistic (NR) limit of the normal contribution coincides with $F_{\rm NR}$. 
(The single dispersion representation for $F$ in $q^2$ is dominated in the NR limit by 
the anomalous cut.)
In the decay case, the situation is different: now, the anomalous cut arises in the double 
spectral representation for $F$, and both the anomalous and the normal pieces are 
of the same order in the NR power counting. Nevertheless, in spite of the complications 
in the decay region related to the 
appearance of the new scale $(\mu-m)^2$, $F_{\rm NR}$ and the NR limit of $F$ are shown 
to be equal to each other.   

\section{\label{sect:ii}Spacelike momentum transfers, equal masses in the loop}
In this section we consider the case of particles of the same mass $m$ 
in the loop and $q^2<0$, but do not restrict the values of $p_1^2$ and $p_2^2$.  

\subsection{Single dispersion representation in $q^2$}
\begin{figure}[bt]
\begin{center}
\begin{tabular}{ccc}
\epsfig{file=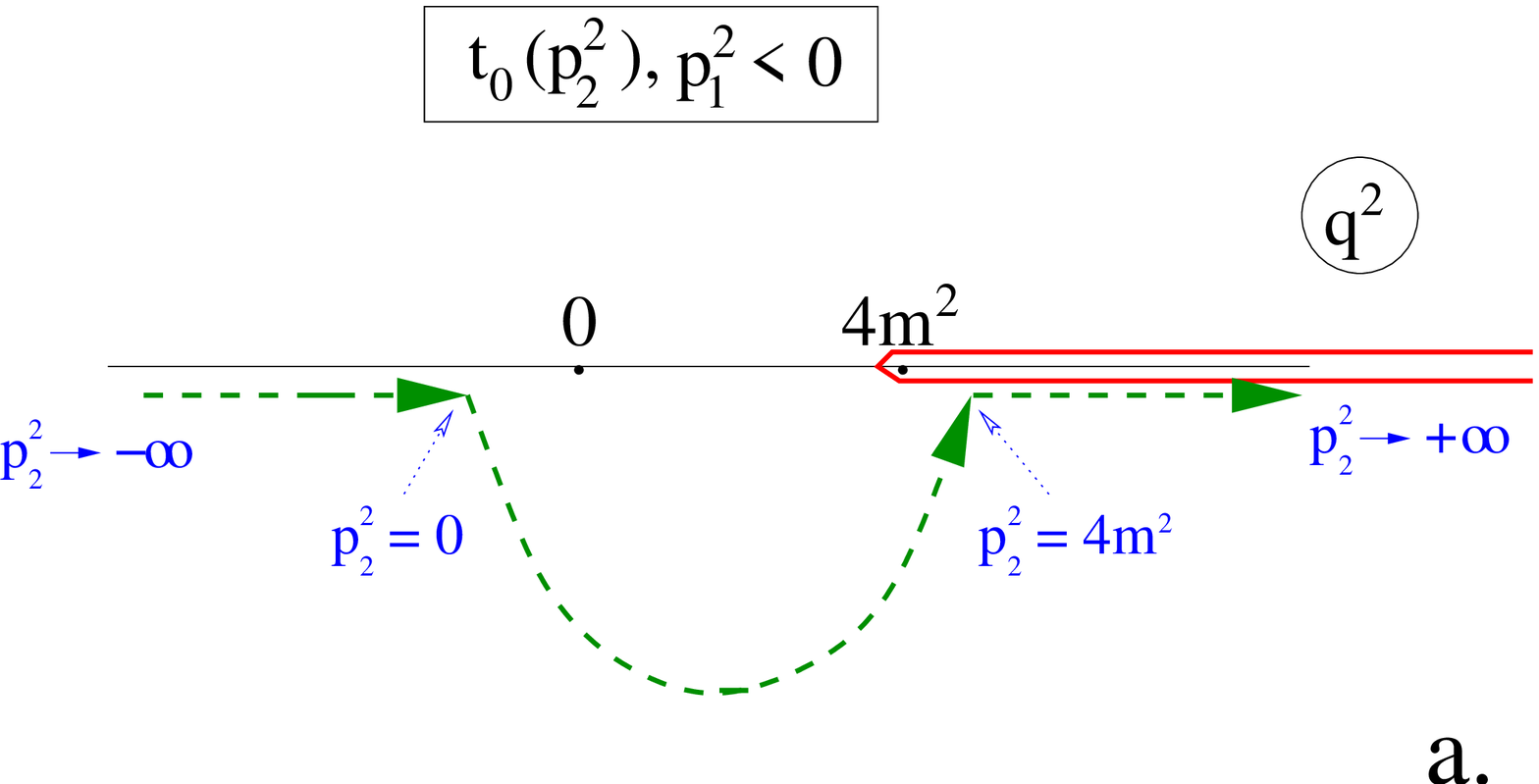,width=5.4cm}\qquad\qquad &
\epsfig{file=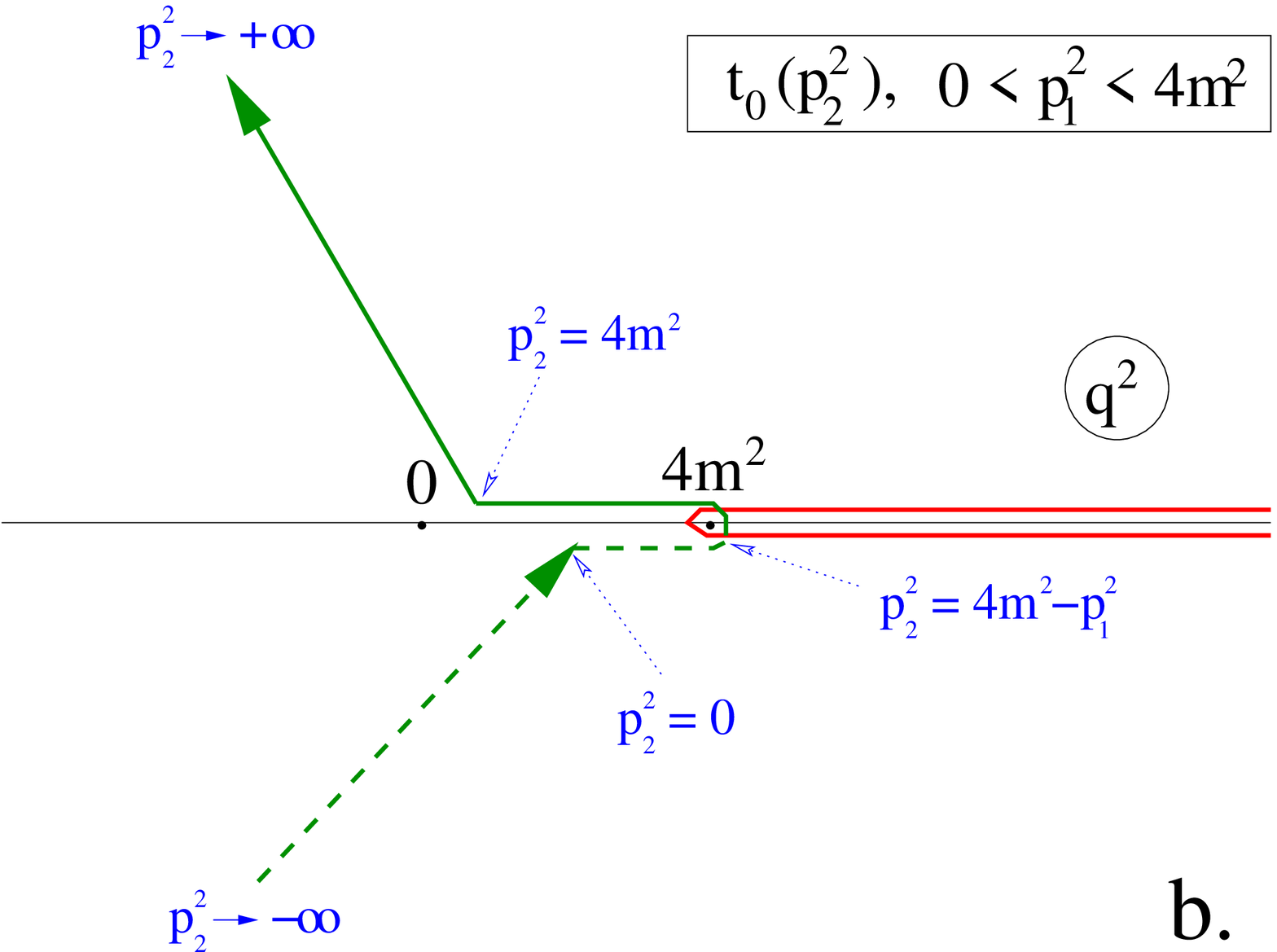,width=5.4cm}\qquad\qquad &
\epsfig{file=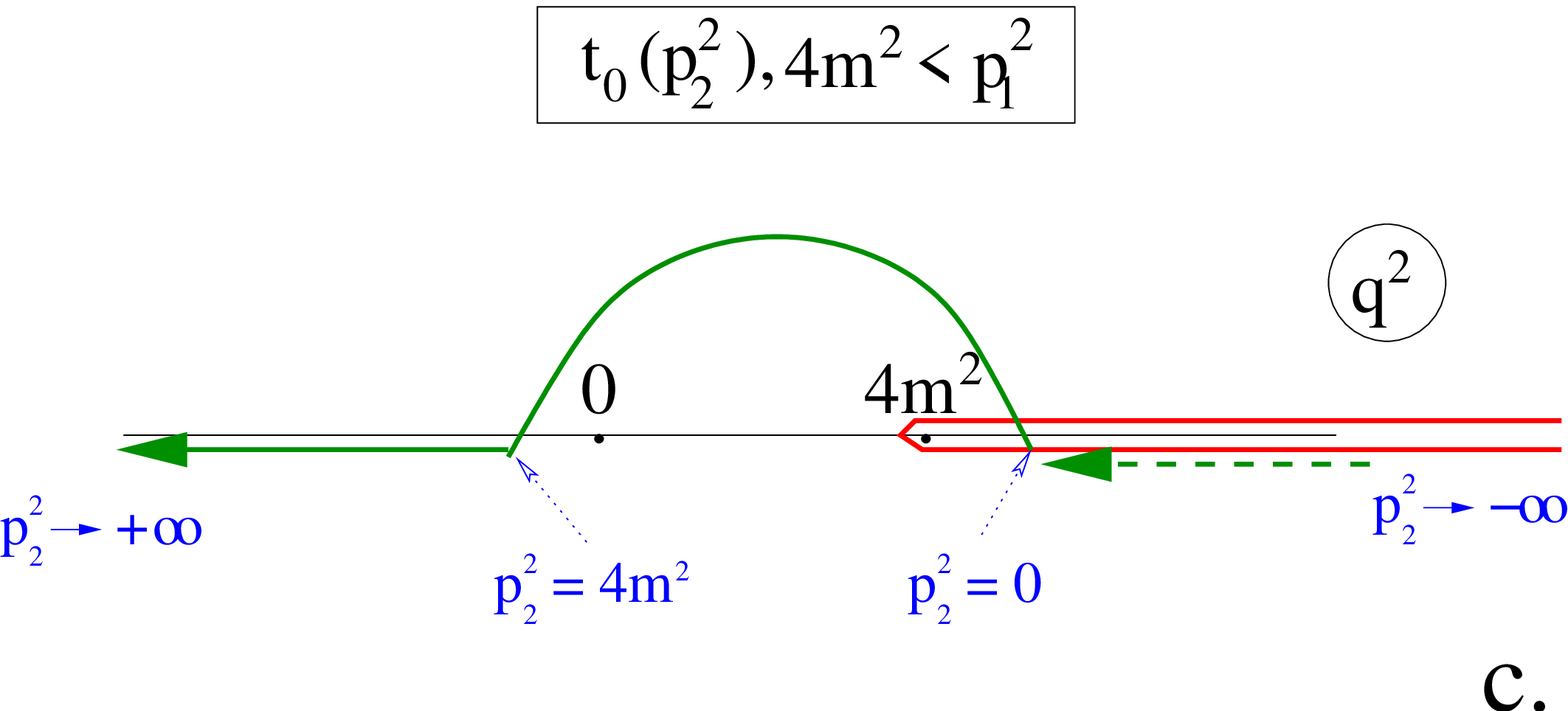,width=5.4cm}
\end{tabular}
\caption{\label{fig:1} 
The trajectory of the branch point $t_0(p_1^2,p_2^2)$ at fixed value of $p_1^2$ (which we denote as $t_0(p_2^2)$) 
in the complex $q^2$-plane: \\
(a): $p_1^2\le 0$, the (dashed-line) trajectory lies on the second sheet, and does not appear on the
physical sheet; 
(b): $0<p_1^2<4m^2$: the trajectory first lies on the second sheet (dashed line), but for 
$p_2^2 > 4m^2-p_1^2$ moves around the normal-cut branch point through the normal cut onto the physical sheet 
(solid line); 
(c): $4m^2 < p_1^2$: similar to case (b). 
The normal cut along the real axis for $q^2>4m^2$ is shown in red.}
\end{center}
\end{figure}

A normal single dispersion representation in $q^2$ may be written as 
\begin{eqnarray}
\label{single}
F(q^2,p_1^2,p_2^2)=\frac{1}{\pi}\int \frac{dt}{t-q^2-i0}\sigma(t,p_1^2,p_2^2).    
\end{eqnarray} 
For $p_1^2<0$ and $p_2^2<0$, the absorptive part $\sigma(t,p_1^2,p_2^2)$ may be calculated by the 
Cutkosky rules, i.e., by placing particles attached to the $q^2$ vertex on the mass shell  
$(m^2-k^2-i0)^{-1}\to 2i\pi\theta(k_0)\delta(m^2-k^2)$. 
The result reads \cite{melikhov}
\begin{eqnarray}
\sigma(t,p_1^2,p_2^2)&=&\frac{1}{16\pi\lambda^{1/2}(t,p_1^2,p_2^2)}
\log\left(
\frac{t-p_1^2-p_2^2+\lambda^{1/2}(t,p_1^2,p_2^2)\sqrt{1-4m^2/t}}
{t-p_1^2-p_2^2-\lambda^{1/2}(t,p_1^2,p_2^2)\sqrt{1-4m^2/t}}\right)\theta(t-4m^2). 
\end{eqnarray}
The function $\sigma(t,p_1^2,p_2^2)$ has the branch point of the logarithm at 
$q^2=t_0(p_1^2,p_2^2)$ given by the solution to the equation 
$(t-p_1^2-p_2^2)^2=\lambda(t,p_1^2,p_2^2)(1-4m^2/t)$, or,
equivalently, to the equation  
\begin{eqnarray}
\frac{p_1^2p_2^2t}{m^2}+\lambda(p_1^2,p_2^2,t)=0.  
\end{eqnarray} 
Explicitly, one finds \cite{karplus,landau}
\begin{eqnarray}
\label{t0}
t^\pm_0= p_1^2+p_2^2-\frac{p_1^2 p_2^2}{2m^2}\pm
\frac{1}{2m^2}\sqrt{p_1^2(p_1^2-4m^2)p_2^2(p_2^2-4m^2)}. 
\end{eqnarray} 
For $p_1^2<0$ or $p_2^2<0$ these branch points lie on the second (unphysical) 
sheet of the function $\sigma$ and do not influence the $q^2$-dispersion representation for $F$.  
However, in the Minkowski region of positive values of $p_1^2$ and $p_2^2$, 
the branch point $t^+_0$, which we hereafter denote simply as $t_0$, may move onto the physical sheet
through the normal cut, thus requiring the modification of the dispersion representation for $F$. 
Let us study the trajectory of the branch point $t_0$ vs. $p_1^2$ and $p_2^2$.
It is a straightforward task, which, however, needs care to guarantee staying at the correct 
branch of the square root, corresponding to the physical values of
$p_1^2$ and $p_2^2$ in the upper complex halfplane. To this end, we introduce the 
variables $\xi_1$ and $\xi_2$ as follows (see \cite{landau_lifshitz}, Eq. (113.11) for details):
\begin{eqnarray}
p_i^2=-m^2\frac{(1-\xi_i)^2}{\xi_i}, \qquad i=1,2.
\end{eqnarray} 
This transformation maps the upper halfplane of the complex variable $p_i^2$ onto the 
internal semicircle with unit radius in the complex $\xi$-plane: 
the region $0<\xi_i<1$ corresponds to $p_i^2<0$, 
the boundary of the semicircle  $\xi_i=\exp(i \varphi_i)$, $0< \varphi_i <\pi$,  
corresponds to the unphysical region $0<p_i^2<4m^2$, 
and the segment $-1<\xi_i<0$ corresponds to $4m^2<p_i^2$. Then 
\begin{eqnarray}
\sqrt{p_i^2(p_i^2-4m^2)}=m^2\frac{1-\xi_i^2}{\xi_i}, 
\end{eqnarray} 
and, for $0<p_i^2<4m^2$, we obtain 
\begin{eqnarray}
\sqrt{p_i^2(p_i^2-4m^2)}=-2i \sin \varphi_i. 
\end{eqnarray} 
We are now ready to study the trajectory of the point $t_0(p_1^2,p_2^2)$ vs. $p_2^2>0$ for a fixed value 
of $p_1^2$ (Fig.~\ref{fig:1}). 
It is convenient to consider three different ranges of $p_1^2$: 
(a) For $p_1^2<0$, the trajectory lies on the second sheet for all values of $p_2^2$, 
and therefore the function is given by its normal dispersion representation in $q^2$. 
(b) For $0<p_1^2<4m^2$, the branch point $t_0$ moves onto the physical sheet through the normal $q^2$-cut if 
$p_2^2$ satisfies the relation $p_1^2+p_2^2>4m^2$. 
(c) For $4m^2<p_1^2$, the situation is similar to the case (b): for $p_2^2>0$ the  
branch point $t_0$ moves onto the physical sheet through the normal $q^2$-cut.
 
Therefore, for external momenta satisfying the relation 
$p_1^2>0$, $p_2^2>0$, $p_1^2+p_2^2>4m^2$, the integration contour in the dispersion representation 
for the form factor depends on the values of $p_1^2$ and $p_2^2$: the contour should be chosen 
such that it embraces both branch points: the normal branch point at $q^2=4m^2$ and the anomalous 
branch point at $q^2=t_0(p_1^2,p_2^2)$. 
\begin{figure}[bt]
\begin{center}
\epsfig{file=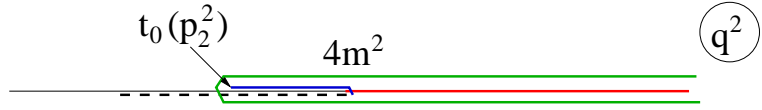,width=10.0cm}
\caption{\label{fig:2} 
The integration contour (green) in the complex $q^2$-plane for 
$0<p_1^2<4m^2$, $0<p_2^2<4m^2$, and $p_1^2+p_2^2>4m^2$: it embraces the anomalous cut (blue), which 
lies along the real axis from $t_0$ to $4m^2$, and the normal cut (red) from $4m^2$ to $+\infty$.}
\end{center}
\end{figure}

Let us consider the single dispersion representation for the form factor in the region 
$0< p_1^2 < 4m^2$, $0 < p_2^2 < 4m^2$, and $4m^2 < p_1^2 +p_2^2$.
This case corresponds to an interesting example of the 
relativistic two-particle bound state, and is necessary for considering the nonrelativistic 
expansion.  

The corresponding $t_0$-trajectory is shown in Fig.~\ref{fig:1}(b). 
Fig.~\ref{fig:2}  gives the integration contour for this case: this contour may be chosen along the real axis from 
$t_0(p_2^2)$ to $+\infty$. It contains two pieces: the normal part from $4m^2$ to $+\infty$
and the anomalous part from $t_0$ to $4m^2$. 

Let us start with the normal part, which has the form  
\begin{eqnarray}
\label{norm}
\sigma_{\rm norm}(t,p_1^2,p_2^2)=
\left\{\begin{array}{ll} 
\frac1{16\pi\sqrt{\lambda(t,p_1^2,p_2^2)}}
\log\left(
\frac{t-p_1^2-p_2^2+\lambda^{1/2}(t,p_1^2,p_2^2)\sqrt{1-4m^2/t}}
{t-p_1^2-p_2^2-\lambda^{1/2}(t,p_1^2,p_2^2)\sqrt{1-4m^2/t}}\right), 
& \quad
(\sqrt{p_1^2}+\sqrt{p_2^2})^2\le t,\\ 
\frac1{8\pi\sqrt{-\lambda(t,p_1^2,p_2^2)}}\arctan
\left(\frac{\sqrt{-\lambda(t,p_1^2,p_2^2)}\sqrt{1-4m^2/t}}{t-p_1^2-p_2^2}\right), 
& 
\quad  
p_1^2+p_2^2\le t\le (\sqrt{p_1^2}+\sqrt{p_2^2})^2,\\ 
\frac1{8\pi\sqrt{-\lambda(t,p_1^2,p_2^2)}}
\left[\pi+\arctan
\left(\frac{\sqrt{-\lambda(t,p_1^2,p_2^2)}\sqrt{1-4m^2/t}}{t-p_1^2-p_2^2}\right)\right], 
& 
\quad  4m^2\le t\le p_1^2+p_2^2.
\end{array}\right. 
\end{eqnarray}
Notice that the normal spectral density does not vanish at the normal threshold $t=4m^2$. 

The discontinuity of the form factor $F(q^2,p_1^2,p_2^2)$ on the anomalous cut is related 
to the discontinuity of the function $\sigma_{\rm norm}(t,p_1^2,p_2^2|m^2)$ and reads 
\begin{eqnarray}
\label{anom}
\sigma_{\rm anom}(t,p_1^2,p_2^2)=
\frac{1}{8\sqrt{-\lambda(t,p_1^2,p_2^2)}},\qquad t_0\le t\le 4m^2.
\end{eqnarray} 
Therefore, the full spectral density has the form 
\begin{eqnarray}
\sigma(t,p_1^2,p_2^2)=
\theta(p_1^2+p_2^2-4m^2)
\theta(t_0\le t\le 4m^2)\sigma_{\rm anom}(t,p_1^2,p_2^2)
+
\theta(4m^2\le t)\sigma_{\rm norm}(t,p_1^2,p_2^2). 
\end{eqnarray}
Clearly, the spectral density given by Eqs. (\ref{norm}) and (\ref{anom}) is a continuous function for 
$t>t_0$.
The spectral representation for the form factor reads  
\begin{eqnarray}
F(q^2,p_1^2,p_2^2)=
\theta(p_1^2+p_2^2-4m^2)
\int\limits_{t_0(p_1^2,p_2^2)}^{4m^2}
\frac{dt}{\pi(t-q^2-i0)}\sigma_{\rm anom}(t,p_1^2,p_2^2)
+
\int\limits_{4m^2}^{\infty}
\frac{dt}{\pi(t-q^2-i0)}\sigma_{\rm norm}(t,p_1^2,p_2^2). 
\end{eqnarray}
For $t_0(p_1^2,p_2^2)<q^2<4m^2$ (in case $p_1^2+p_2^2>4m^2$)
the imaginary part of the form factor comes from the anomalous
part, while for $q^2>4m^2$ it comes from the normal part.

\subsection{Double dispersion representation in $p_1^2$ and $p_2^2$}
For $q^2\le 0$, 
the triangle diagram may be written 
as the double dispersion representation
\begin{eqnarray}
\label{double}
F(q^2,p_1^2,p_2^2)=\int \frac{ds_1}{\pi(s_1-p_1^2-i0)}\frac{ds_2}{\pi(s_2-p_2^2-i0)}
\Delta(q^2,s_1,s_2). 
\end{eqnarray} 
The double spectral density $\Delta(q^2,s_1,s_2)$ may be obtained by placing all particles 
in the loop on the mass shell and taking the off-shell external momenta 
$p_1\to \tilde p_1$, $p_2\to \tilde p_2$, 
such that $\tilde p_1^2=s_1$, $\tilde p_2^2=s_2$, and 
$(\tilde p_1-\tilde p_2)^2=q^2$ is fixed \cite{anisovich}: 
\begin{eqnarray}
\Delta(q^2,s_1,s_2)&=&\frac1{8\pi}
\int dk_1 dk_2 dk_3 \delta(\tilde p_1-k_2-k_3)\delta(\tilde p_2-k_3-k_1)
\theta(k_1^0)\delta(k^2_1-m^2)\theta(k_2^0)\delta(k^2_2-m^2)\theta(k_3^0)\delta(k^3_2-m^2), \nonumber \\
&&
\quad \tilde p_1^2=s_1, \quad \tilde p_2^2=s_2, \quad (\tilde p_1-\tilde p_2)^2=q^2. 
\end{eqnarray}
Explicitly, one finds 
\begin{eqnarray}
\label{thetafunction}
\Delta(q^2,s_1,s_2)=\frac{1}{16\lambda^{1/2}(s_1,s_2,q^2)}
\theta\left({s_1-4m^2}\right)
\theta\left({s_2-4m^2}\right)
\theta\left[{\left(q^2(s_1+s_2-q^2)\right)^2-\lambda(s_1,s_2,q^2)\lambda(q^2,m^2,m^2)}\right]. 
\end{eqnarray}
The solution of the $\theta$-function gives the following allowed intervals 
for the integration variables $s_1$ and $s_2$: 
\begin{eqnarray}
\label{limits}
4m^2<&s_2,&
\nonumber\\
s_1^-(s_2,q^2)<&s_1&<s_1^+(s_2,q^2), 
\end{eqnarray}
where  
\begin{eqnarray}
\label{s1pm_equal_masses}
s_1^\pm(s_2,q^2)&=&s_2+q^2-\frac{s_2 q^2}{2 m^2}
\pm\frac{\sqrt{s_2(s_2-4m^2)}\sqrt{q^2(q^2-4m^2)}}{2m^2}.
\end{eqnarray}
The final double dispersion representation for the triangle diagram at $q^2<0$ takes the form\footnote{
The easiest way to obtain this double dispersion representation 
is to introduce light-cone variables in the Feynman expression, and to
choose the 
reference frame where $q_+=0$ (which restricts $q^2$ to $q^2<0$). Then the $k_-$ integral is easily done,
and the remaining $y$ and $k_\perp$ integrals may be written in the
form (\ref{double});  
details can be found in \cite{anisovich}.}
\begin{equation}
\label{fftrans}
F(q^2,p_1^2,p_2^2)=\int\limits^\infty_{4m^2}\frac{ds_2}{\pi(s_2-p_2^2-i0)}
\int\limits^{s_1^+(s_2,q^2)}_{s_1^-(s_2,q^2)}\frac{ds_1}{\pi(s_1-p_1^2-i0)}
\frac{1}{16\lambda^{1/2}(s_1,s_2,q^2)}.
\end{equation}
Notice the relation $s_1^-(s_2,q^2)>4m^2$, which holds for all $s_2>4m^2$ at $q^2<0$: this guarantees 
that the integration region in $s_1$ always remains above the normal threshold. 
Clearly, the integration region does not depend on the values of $p_1^2$ and $p_2^2$. 
Essential for us is that no anomalous cuts emerge in the double dispersion 
representation in $p_1^2$ and $p_2^2$ for $q^2<0$. This makes the double dispersion representation particulary 
convenient for treating the triangle diagram for values of $p_1^2$ and $p_2^2$ above 
the thresholds. 
One should just take care about the appearance of the absorptive parts. 

\section{\label{sect:iii}Double spectral representation for the decay kinematics}
Now we discuss the triangle diagram with particles of different masses in the loop, $m<\mu$, 
and consider the decay kinematics $0<q^2<(\mu-m)^2$. 
We have in mind the application to processes corresponding to the overthreshold 
values $p_1^2>(\mu+m)^2$ and $p_2^2>4m^2$, such as, e.g., the $K\to 3\pi$ decay \cite{k3pi}. 
As we have seen in the previous section, the single dispersion representation in $q^2$ 
is rather complicated for $p_1^2$ and $p_2^2$ above the two-particle thresholds 
already for equal masses in the loop. 
The situation is much worse for unequal masses in the loop. 
On the other hand, we shall see that the double spectral representation in $p_1^2$ and 
$p_2^2$ is rather simple in this case for $q^2<(\mu-m)^2$. 
We start from the region $q^2<0$, where the double dispersion representation has the 
standard form both for equal and unequal masses in the loop. 
We then perform the analytic continuation in
$q^2$ and observe the appearance of the anomalous contribution in the double spectral representation. 

\subsection{Transition form factor at $q^2<0$}
For $q^2<0$, the double dispersion representation has a form very similar to the case of 
equal masses \cite{melikhov}: 
\begin{equation}
\label{fftrans2}
F(q^2,p_1^2,p_2^2)=\int\limits^\infty_{4m^2}\frac{ds_2}{\pi(s_2-p_2^2)}
\int\limits^{s_1^+(s_2,q^2)}_{s_1^-(s_2,q^2)}\frac{ds_1}{\pi(s_1-p_1^2)}
\frac{1}{16\lambda^{1/2}(s_1,s_2,q^2)}, 
\end{equation}
where  
\begin{eqnarray}
\label{s1pm}
s_1^\pm(s_2,q^2)&=&
\frac{s_2(m^2+\mu^2-q^2)+2 m^2 q^2}{2m^2}
\pm\frac{\lambda^{1/2}(s_2,m^2,m^2)\lambda^{1/2}(q^2,\mu^2,m^2)}{2m^2}. 
\end{eqnarray}
A new feature compared with the case of equal masses in the loop  
is the appearance of the region $0<q^2<(\mu-m)^2$, which was absent in the equal-mass case.  
This region corresponds to the decay of a particle of mass $\mu$ to
a particle of mass $m$ with the emission of a particle of mass $\sqrt{q^2}$.

\subsection{Transition form factors at $q^2>0$}
The form factor in the region $0<q^2<(\mu-m)^2$ may be obtained by analytic 
continuation of the expression (\ref{fftrans}). 
Let us consider the structure of the singularities of the integrand in Eq.~(\ref{fftrans2}) 
in the complex $s_1$-plane for a fixed 
real value of $s_2$ in the interval $s_2>4m^2$. 
\begin{figure}
\begin{center}
\mbox{\epsfig{file=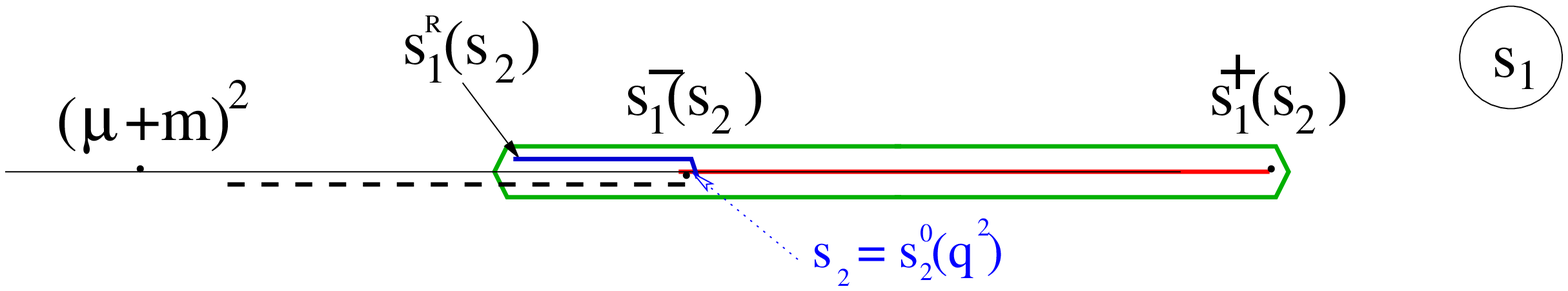,width=8cm}}
\end{center}
\caption{\label{fig:double}
Singularities of the function $\Delta(q^2,s_1,s_2)$ in the complex $s_1$
plane as a function of $s_2$ for $q^2>0$. (This corresponds to an external $s_2$ 
and an internal $s_1$ integration). 
The trajectory $s_1^R(s_2)$ at fixed
$q^2>0$ is shown: for $s_2<s_2^0$ the branch point $s_1^R(s_2)$ 
remains on the unphysical sheet (dashed line), 
but, as soon as $s_2>s_2^0$, it goes onto the physical sheet and moves to the
left from the left boundary of the normal cut $s_1^-$.
Respectively, for $s_2<s_2^0$ the integration contour in the complex 
$s_1$-plane 
may be chosen along the interval $[s_1^-,s_1^+]$. For 
$s_2>s_2^0$, however, the contour should embrace the point $s_1^R$, and
therefore the inegration contour contains two segments: the ``anomalous'' segment 
from $s_1^R$ to $s_1^-$, and the ``normal'' segment from $s_1^-$ to $s_1^+$.}
\end{figure}

\begin{figure}
\begin{center}
\mbox{\epsfig{file=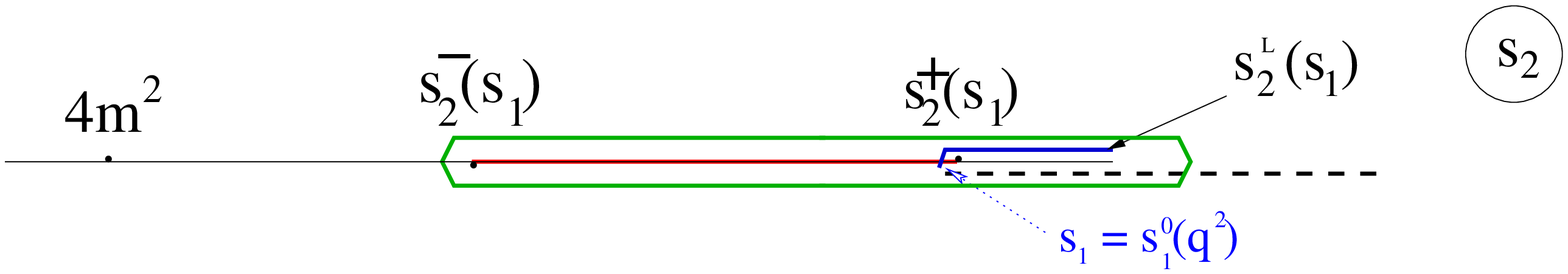,width=8cm}}
\end{center}
\caption{\label{fig:double_b}
Singularities of the function $\Delta(q^2,s_1,s_2)$ in the complex $s_2$
plane as a function of $s_1$ for $q^2>0$. (This corresponds to an external $s_1$  
and an internal $s_2$ integration). 
The trajectory $s_2^L(s_1)$ at fixed
$q^2>0$ is shown: for $s_1<s_1^0$ the branch point $s_2^L(s_1)$ 
remains on the unphysical sheet (dashed line), 
but, as soon as $s_1>s_1^0$, it goes onto the physical sheet and moves to the
right from the right boundary of the normal cut $s_2^+$. The notations are self-evident. 
}
\end{figure}

The integrand has singularities (branch points) 
related to the zeros of the function $\lambda(s_1,s_2,q^2)$ at $s_1^L=(\sqrt{s_2}-\sqrt{q^2})^2$ and 
$s_1^R=(\sqrt{s_2}+\sqrt{q^2})^2$. As $q^2\le 0$, these singularities lie on the unphysical sheet. 
However, as $q^2$ becomes positive, the point $s_1^R$ may move onto the physical sheet through the cut 
from $s_1^-$ to $s_1^+$. This happens for values of the variable $s_2>s_2^0$, with 
$s_2^0$ obtained as the solution to the equation $s_1^R(s_2,q^2)=s_1^-(s_2,q^2)$. Explicitly, 
one finds  
\begin{eqnarray}
\sqrt{s_2^0}=\frac{\mu^2-m^2-q^2}{\sqrt{q^2}}.
\end{eqnarray}
The trajectory of the point $s_1^R(s_2,q^2)$ in the complex $s_1$-plane at fixed $q^2>0$ vs. $s_2$ 
is shown in
Fig.~\ref{fig:double}. 
As $q^2>0$, for $s_2>s_2^0(q^2)$ the
integration contour in the complex $s_1$-plane should be deformed such that it embraces the 
points $s_1^R$ and
$s_1^+$. Respectively, the $s_1$-integration contour contains the two segments: the normal part from 
$s_1^-$ to $s_1^+$, and the anomalous part from $s_1^R$ to $s_1^-$. The double spectral density 
for the anomalous piece is just the discontinuity of the function 
$1/\sqrt{\lambda(s_1,s_2,q^2)}$. It can be easily calculated as follows: 
Recall the relation $\sqrt{\lambda(s_1,s_2,q^2)}=\sqrt{s_1-s_1^L}\sqrt{s_1-s_1^R}$. The branch point 
$s_1^L$ lies on the unphysical sheet, therefore the function $\sqrt{s_1-s_1^L}$ is continuous 
on the anomalous cut located on the physical sheet. Thus we have to calculate the 
discontinuity of the
function $1/\sqrt{s_1-s_1^R}$ which is just twice the function itself. As the result, the discontinuity
of the function $1/\sqrt{\lambda(s_1,s_2,q^2)}$ on the anomalous cut is just 
$2/\sqrt{\lambda(s_1,s_2,q^2)}$.
Finally, the full double spectral density including the normal and the anomalous pieces 
takes the form\footnote{In \cite{braun,melikhov} the double spectral density was obtained by a rather
complicated procedure, considering first the single spectral representation in $s_2$. We point out that
this step is unnecessary: the final result may be obtained just starting from the double spectral 
representation at $q^2<0$, where only the normal contribution is present. The derivation applied here 
promises strong simplifications for obtaining the double spectral representation in the production region
$q^2>(\mu + m)^2$.} 
\begin{eqnarray}
\label{4deltas}
\Delta(q^2,s_1,s_2|\mu,m,m)&=&
\frac{\theta(s_2-4m^2)\theta(s_1^-<s_1<s_1^+)}{16\lambda^{1/2}(s_1,s_2,q^2)}
+\frac{2\theta(q^2)\theta(s_2-s_2^0)\theta(s_1^R<s_1<s_1^-)}{16\lambda^{1/2}(s_1,s_2,q^2)}.
\end{eqnarray}
The first term in (\ref{4deltas}) relates to the Landau-type contribution
emerging when all
intermediate particles go on mass shell, while the second term describes the
anomalous contribution.

The result (\ref{4deltas}) for $\Delta$ holds for $\mu > m$ implying the ``external'' 
$s_2$-integration, and the ``internal'' $s_1$-integration. The location of the integration 
region for this case is shown in Fig.~\ref{fig:double}. Fig.~\ref{fig:double_b} gives the 
integration contour in the complex $s_2$ plane for the opposite order of the integrations. 

\begin{figure}[t]
\begin{center}
\epsfig{file=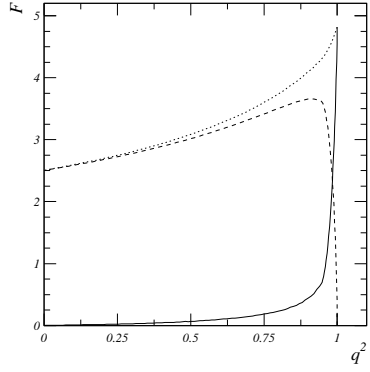,height=5.0cm}
\caption{\label{fig:plot} 
A typical behavior of the function 
$F(q^2,p_1^2,p_2^2)$ vs. $q^2$ for $0<q^2<(\mu-m)^2$ at
fixed $p_1^2$ and $p_2^2$. 
The parameters are chosen such that $(\mu-m)^2=1$ GeV$^2$. 
Dashed: normal part, solid: anomalous part, dotted: full function (sum of both parts). }
\end{center}
\end{figure}

The final representation for the form factors at
$0<q^2<(\mu-m)^2$
takes the form  
\begin{eqnarray}
\label{final1}
F(q^2,p_1^2,p_2^2)&=&
\int\limits_{4m^2}^\infty\frac{ds_2}{\pi(s_2-p_2^2-i0)}
\int\limits_{s_1^-(s_2,q^2)}^{s_1^+(s_2,q^2)}
\frac{ds_1}{\pi(s_1-p_1^2)}
\frac{1}{16\lambda^{1/2}(s_1,s_2,q^2)}
\nonumber\\
&+&
2\theta\left(0<q^2<(\mu-m)^2\right)\int\limits_{s_2^0(q^2)}^\infty\frac{ds_2 }{\pi(s_2-p_2^2-i0)}
\int\limits_{s_1^R(s_2,q^2)}^{s_1^-(s_2,q^2)}
\frac{ds_1}{\pi(s_1-p_1^2)}\frac{1}{16\lambda^{1/2}(s_1,s_2,q^2)}.
\end{eqnarray}
A typical behavior of the anomalous and the normal contributions is plotted in Fig.~\ref{fig:plot}:
the normal contribution first rises at small values of $q^2$ but then drops down steeply and 
vanishes at zero recoil. The anomalous contribution is zero at $q^2=0$, remains small at small 
$q^2>0$, but rises steeply near zero recoil, providing a smooth behavior of the full form factor. 

\begin{figure}[h]
\begin{center}
\epsfig{file=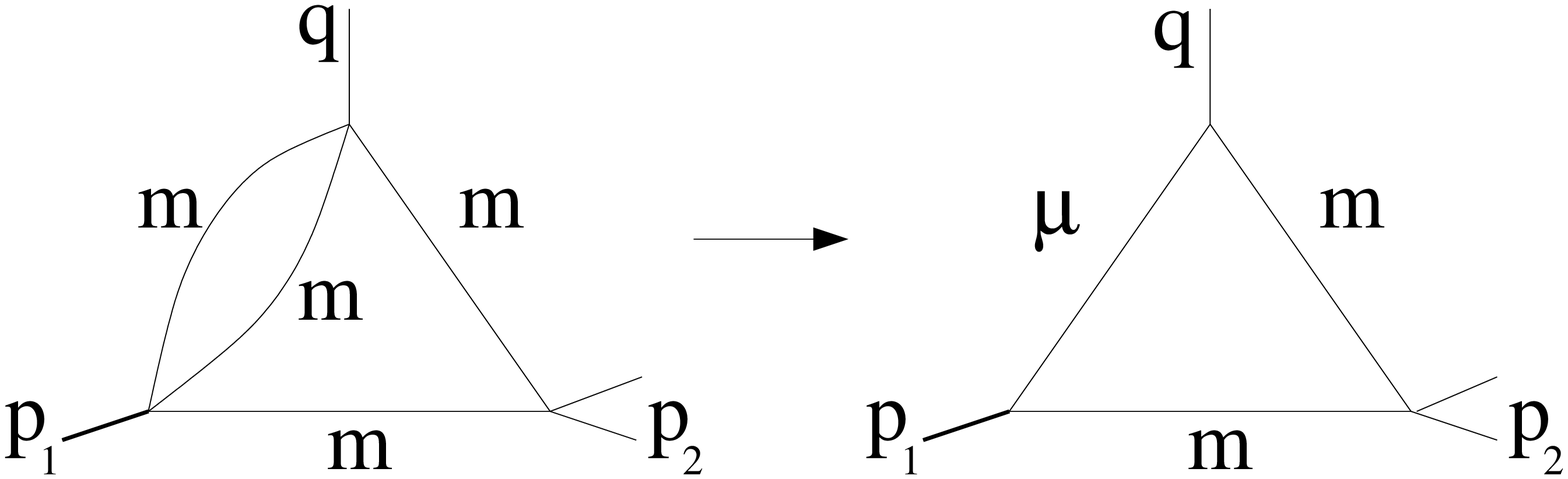,width=9.0cm}
\caption{\label{fig:k3pi} 
The triangle-diagram contribution to the $K\to 3\pi$ amplitude may be reduced to the 
integral over $\mu^2$ of the diagram $F$. }
\end{center}
\end{figure}

We point out that the representation (\ref{final1}) is particularly suitable for application to processes 
where $p_1^2$ and $p_2^2$ are above two-particle thresholds: in this case the single spectral
representation in $q^2$ becomes extremely complicated, with a nontrivial integration contour in the
complex $q^2$-plane, whereas the double dispersion representation in $p_1^2$ and $p_2^2$ has the  
simple form given above. For values of $p_1^2$ and $p_2^2$ above the thresholds one just 
has to take into account the appearance of the absorptive parts in the $s_1$ and $s_2$ integrals. 
A possible application of this representation may be the calculation of the triangle-diagram
contribution to the three-body decay \cite{anisovich_anselm}, e.g., to the $K\to 3\pi$ decay 
\cite{k3pi}, Fig.~\ref{fig:k3pi}. 
In this case the diagram with the pion loop may be represented as the $\mu^2$ integral of the 
triangle diagram considered here, and one obtains the expression for the 
values $p_1^2=M_K^2>9m_\pi^2$, $p_2^2>4m_\pi^2$, and $q^2=m_\pi^2$. 
The emerging absorptive parts may then be easily calculated from the double spectral representation. 
The problem would be technically very involved if one uses the single spectral representation 
in $q^2$, as can be seen from the complicated structure of the integration contour in 
Section \ref{sect:ii}. 

\section{\label{sect:iv} The nonrelativistic expansion for the case of decay kinematics}
In this section we perform the nonrelativistic expansion of the double spectral representation for the 
triangle diagram $F$ for the case $\mu > m$ and compare it with the triangle diagram of the 
nonrelativistic field theory $F_{\rm NR}$. 
For the latter we also obtain a double spectral representation. However, 
the double spectral representations for 
$F$ and $F_{\rm NR}$ have rather different properties. Nevertheless, the  
two expressions are shown to match to each other.  

\subsection{Nonrelativistic expansion of the relativistic triangle diagram}
Let us look at the behavior of the anomalous and the normal
contributions in the NR limit. To this end, we introduce new variables: 
instead of $p_1^2=M_1^2$, we use $M_1=\mu+m-\epsilon_1$, 
instead of $p_2^2=M_2^2$, we use $M_2=2m-\epsilon_2$, and the NR approximation requires 
$\epsilon_i\ll m, \mu$. Instead of $q^2$, we use the variable $u$ defined by 
\begin{eqnarray}
q^2=(\mu-m)^2-2m(\mu+m)u^2.    
\end{eqnarray}
The maximal decay momentum transfer $q^2=(\mu-m)^2$ corresponds to $u=0$, and the decay region 
is $u>0$. The meaning of the coefficient $2m(\mu+m)$ will be clear from comparison with the NR 
field theory.  
The consistency of the NR approximation requires the momentum transfer 
to be limited, therefore 
\begin{eqnarray}
u^2\le O(\Lambda/m), 
\end{eqnarray}
where $\Lambda$ is a constant which does not scale with the mass. 
In the NR limit, one finds
\begin{eqnarray}
\label{fnorm}
F_{\rm norm}(u,\epsilon_1,\epsilon_2)&=&\frac{1}{64\pi^2\sqrt{m(\mu+m)(\mu-m)}}
\int_{0}^\infty
\frac{dz_2}{(z_2+\epsilon_2)}
\int_{z_1^-}^{z_1^+}
\frac{dz_1}{(z_1+\epsilon_1)\sqrt{z_1-z_1^R}},\\ 
\label{fanom}
F_{\rm anom}(u,\epsilon_1,\epsilon_2)&=&\frac{1}{32\pi^2\sqrt{m(\mu+m)(\mu-m)}}
\int_{z_2^0}^\infty
\frac{dz_2}{(z_2+\epsilon_2)}
\int_{z_1^R}^{z_1^-}
\frac{dz_1}{(z_1+\epsilon_1)\sqrt{z_1-z_1^R}}.
\end{eqnarray}
The condition $\theta(0<q^2<(\mu-m)^2)$, which defines the region where the anomalous contribution 
is nonvanishing, takes the form $z_2\le m$. The latter condition is automatically fulfilled 
in the NR limit which requires $z_{1,2}\ll m$.  

The integration limits  
\begin{eqnarray} 
\label{limits1}
z_2^0=\frac{2\mu m(\mu+m)}{(\mu-m)^2}u^2,\qquad
z_1^R=z_2-\frac{2\mu}{(\mu-m)}u^2, \qquad
z_1^{\pm}=\frac{2\mu}{(\mu+m)}\left(\sqrt{z_2}\pm u\sqrt\frac{m(\mu+m)}{2\mu}\right)^2 
\end{eqnarray} 
are obtained by keeping the leading nonrelativistic terms of the values 
$s_2^0$, $s_1^R$, and $s_1^{\pm}$, respectively. 
Interestingly, both the normal and the anomalous contributions 
remain finite in the NR limit.\footnote{We regard this to be rather unexpected for the following reason:  
The appearance of the anomalous contribution is related to the cumbersome migration of singularities 
in the complex plane from the unphysical sheet onto the physical sheet through the normal cut. 
In the double dispersion representation for the triangle diagram of
the NR field theory this does not occur, 
and the double dispersion representation for the NR triangle diagram has no anomalous contribution. 
Therefore, one might expect that also in the double dispersion representation for the 
relativistic triangle diagram only the normal contribution survives in the NR limit. 
Here we see that this is not the case: both the normal and the anomalous contributions
survive. 
In Section~\ref{sect:iv.b} we see that the same expression emerges as the normal 
contribution of the NR double dispersion representation.} 
Moreover, in the NR limit the normal part contains only the odd powers of $u$, whereas the  
terms of odd powers in $u$ cancel in the sum of the normal and the anomalous parts. 
Therefore, the only role of $F_{\rm norm}$ in the case of the decay kinematics is to cancel the terms of the odd powers 
in $u$. Recall that this is completely different from the case of the elastic kinematics: in the latter case 
the anomalous part is absent at all. 

It is convenient to obtain the form factor as an expansion in powers of $u$ and $h$, where $h=(\mu-m)/(\mu+m)$. 
For the sum of the normal and the anomalous parts, we find
\begin{eqnarray}
\label{30}
F(u^2,\epsilon_1,\epsilon_2,)&=&\frac{1}{32\pi{m}^{3/2}\left(\sqrt{\epsilon_1}+\sqrt{\epsilon_2}\right)}
-h \frac{7\sqrt{\epsilon_1}+8\sqrt{\epsilon_2}}{192\pi{m}^{3/2}\left(\sqrt{\epsilon_1}+\sqrt{\epsilon_2}\right)^2}
\nonumber\\
&+&u^2\left[-\frac{1}{96\pi\sqrt{m}\left(\sqrt{\epsilon_1}+\sqrt{\epsilon_2}\right)^3}
+h\frac{13\sqrt{\epsilon_1}+22\sqrt{\epsilon_2}}{960\pi\sqrt{m}\left(\sqrt{\epsilon_1}+\sqrt{\epsilon_2}\right)^4}
\right] 
+\cdots
\end{eqnarray}

\subsection{\label{sect:iv.b}Triangle diagram in nonrelativistic field theory}
Let us first set up the nonrelativistic kinematics:
\begin{eqnarray}
p_1^0=M_1+\frac{\vec p_1^2}{2M_1}=\mu+m-\epsilon_1+\frac{\vec p_1^2}{2(\mu+m)}, 
\qquad
p_2^0=M_2+\frac{\vec p_2^2}{2M_2}=2m-\epsilon_2+\frac{\vec p_2^2}{4m},
\end{eqnarray}
where we have neglected terms of order $O((\vec p^2)^2/m^4)$ and $O(\vec p^2 \epsilon/m^3)$. 
We now calculate the 4-momentum transfer 
$q^2=(p_1-p_2)^2\simeq (M_1-M_2)^2-M_1M_2 (\vec v_1-\vec v_2)^2$, 
where $\vec v_i=\vec p_i/M_i$, $i=1,2$. Thus, the NR form factor depends on the 
square of the three-dimensional velocity transfer $v^2\equiv \vec v^2$, $\vec v\equiv \vec v_1-\vec v_2$, which 
is reduced to $\vec q^2$ only in the elastic case $M_1=M_2$. 

The propagator of a NR particle has the form $D^{-1}_c(E,\vec k)=-2mE+\vec k^2-i0$ \cite{anisovich_book}, and 
the NR triangle diagram reads 
\begin{eqnarray}
\label{Fnr1}
F_{\rm NR}(\vec v^2,\epsilon_1,\epsilon_2)&=&
\frac{1}{(2\pi)^4 i}
\int 
\frac{dE d^3 k}{
(-2mE+\vec k^2-i0)
(-2\mu(E_1-E)+(\vec p_1-\vec k)^2-i0)
(-2m(E_2-E)+(\vec p_2-\vec k)^2-i0)}, \nonumber\\
&& E_1=\epsilon_1+\frac{\vec p_1^2}{2(\mu+m)}, \quad
E_2=\epsilon_2+\frac{\vec p_2^2}{4m}.
\end{eqnarray}
The $E$-integration is easily performed by closing the integration contour in the lower
complex semiplane. Introducing the new integration variable $\vec w=\vec k/m$ 
(the velocity of the spectator particle in the diagram), we obtain
\begin{eqnarray}
\label{Fnr2}
F_{\rm NR}(\vec v^2,\epsilon_1,\epsilon_2)&=&
\frac{m}{64\pi^3\mu}
\int 
\frac{d^3 w}{
\left[\frac{m(\mu+m)}{2\mu}(\vec v_1-\vec w)^2+\epsilon_1-i0\right]
\left[m(\vec v_2-\vec w)^2+\epsilon_2-i0\right]}. 
\end{eqnarray}
The last equation may be written in the form of the double spectral representation 
\begin{eqnarray}
\label{Fnr3}
F_{\rm NR}(\vec v^2,\epsilon_1,\epsilon_2)&=&
\int \frac{dz_1}{\pi(z_1+\epsilon_1-i0)} \frac{dz_2}{\pi(z_2+\epsilon_2-i0)}
\Delta_{\rm NR}(z_1,z_2,\vec v^2)
\end{eqnarray}
with
\begin{eqnarray}
\label{Fnr4}
\Delta_{\rm NR}(z_1,z_2,(\vec v_1-\vec v_2)^2)=
\frac{m}{64\pi \mu}
\int d^3 w
\delta\left(\frac{m(\mu+m)}{2\mu}(\vec w -\vec v_1)^2-z_1 \right)
\delta\left(m(\vec w -\vec v_2)^2-z_2 \right).
\end{eqnarray}
Performing the integration over $d^3 w$, we arrive at the following double spectral representation 
of the NR field 
theory:\footnote{This expression looks very much like $F_{\rm norm}$ but is in fact different: 
compared with (\ref{Fnr}), the denominator of (\ref{fnorm}) contains the 
term $z_1-z_2$ which cannot be neglected; moreover, the limits of the $z_1$ 
integration are different.} 
\begin{eqnarray}
\label{Fnr}
F_{\rm NR}(v^2,\epsilon_1,\epsilon_2)&=&\frac{1}{64 \pi^2 m(\mu+m)}
\int_0^\infty \frac{dz_2}{z_2+\epsilon_2}
\int_{\tilde z_1^-}^{\tilde z_1^+}\frac{dz_1}{z_1+\epsilon_1}\frac1v,\qquad
\tilde z_1^\pm=\frac{(\mu+m)}{2\mu}
\left(\sqrt{z_2}\pm v\sqrt{m}\right)^2.
\end{eqnarray}
The NR form factor may now be obtained in analytic form as the expansion in powers of $v^2$
\begin{eqnarray}
F_{\rm NR}(v^2,\epsilon_1,\epsilon_2)=\frac{1}{16\pi\sqrt{m}(\mu+m)\left(\sqrt{\epsilon_1}+
\sqrt{\frac{2\mu}{\mu+m}}\sqrt{\epsilon_2}\right)}
-
\frac{\sqrt{m} }{48\pi(\mu+m)\left(\sqrt{\epsilon_1}+
\sqrt{\frac{2\mu}{\mu+m}}\sqrt{\epsilon_2}\right)^3}v^2+O(v^4). 
\end{eqnarray}
Finally, we may expand this expression in powers of $h$: 
\begin{eqnarray}
F_{\rm NR}(v^2,\epsilon_1,\epsilon_2)&=&\frac{1}{32\pi{m}^{3/2}\left(\sqrt{\epsilon_1}+\sqrt{\epsilon_2}\right)}
-h \frac{3\sqrt{\epsilon_1}+2\sqrt{\epsilon_2}}{64\pi{m}^{3/2}\left(\sqrt{\epsilon_1}+\sqrt{\epsilon_2}\right)^2}
\nonumber\\
&+&v^2\left[-\frac{1}{96\pi\sqrt{m}\left(\sqrt{\epsilon_1}+\sqrt{\epsilon_2}\right)^3}
+h\frac{5\sqrt{\epsilon_1}+2\sqrt{\epsilon_2}}{192\pi\sqrt{m}\left(\sqrt{\epsilon_1}+\sqrt{\epsilon_2}\right)^4}
\right] 
+\cdots
\end{eqnarray}
where $\cdots$ denote terms of higher orders in $h$ and $v^2$. 
For comparison with $F(u^2,\epsilon_1,\epsilon_2)$, Eq.~(\ref{30}), one should take into account that the variables $u^2$
and $v^2$ differ from each other. Their relationship is obtained from the equation  
$q^2=(M_1-M_2)^2-M_1M_2 v^2=(\mu-m)^2-2m (\mu+m) u^2$, which gives, to the necessary NR accuracy, 
\begin{eqnarray}
v^2=u^2-h\frac{\epsilon_1-\epsilon_2}{m}.
\end{eqnarray}
Making use of this relation, $F_{\rm NR}(v^2,\epsilon_1,\epsilon_2)$ and the NR expansion of 
$F(u^2,\epsilon_1,\epsilon_2)$ 
perfectly match each other.

\section{Summary and Conclusions} 
We have presented a detailed analysis of dispersion representations for the triangle diagram, laying main
emphasis on the appearance of the anomalous contributions to these representations. 
In some kinematic regions the properties of the triangle
diagram and the amplitudes of the corresponding processes are mainly determined by 
the anomalous contributions. A message we would like to convey to the reader
is that in many cases the double spectral representations in $p_1^2$ and $p_2^2$ provide great technical 
advantages compared to the use of the single representation in $q^2$. This is clearly the case for 
$p_1^2$ and $p_2^2$ above the thresholds and $q^2$ in the decay region $0<q^2<(\mu-m)^2$. 
Several actual physical problems belong to this class of problems.  

The results presented in this paper are summarized below: 

1. We pointed out that at spacelike momentum tranfer $q$, $q^2<0$, and for any values of $p_1^2$ and $p_2^2$, 
the double dispersion representation in $p_1^2$ and $p_2^2$ is particularly simple and contains 
only the normal cut. The calculation of the triangle diagram in this case may be easily done for all 
values of $p_1^2$ and $p_2^2$, including the values above the thresholds and complex values. In the same situation, 
the single spectral representation in $q^2$ contains, in addition, the anomalous cut, making the 
application of the single dispersion representation a very involved problem. 

2. For the decay kinematics $0<q^2<(\mu-m)^2$, we presented a new derivation of the anomalous contribution 
to the double spectral representation. The presented approach allows an extension of double spectral representations
also to higher momentum tranfers $q^2>(\mu+m)^2$. 

In the decay region $0<q^2<(\mu-m)^2$, the double spectral 
representation in $p_1^2$ and $p_2^2$ is shown to provide a very convenient  
tool for considering processes at $p_1^2$ and $p_2^2$ above the thresholds. 
The application of the single spectral representation in $q^2$ faces in this 
case severe technical problems. 

3. We analysed the double spectral representation of the triangle diagram in the region near the
thresholds in $p_1^2$ and $p_2^2$ and for $q^2\simeq (\mu-m)^2$, where the nonrelativistic expansion is possible. 
We have shown that in this case both the normal and the anomalous contributions in $F$ are of the same order
in the nonrelativistic power counting. We also constructed the double dispersion representation of the triangle 
diagram of the nonrelativistic field theory, $F_{\rm NR}$, and demonstrated that this representation does not contain 
the anomalous contribution. Nevertheless, in spite of the complications in the decay region related to the 
appearance of the new scale $(\mu-m)^2$, the $F_{\rm NR}$ and the nonrelativistic limit of $F$ are shown to match each other.    

\acknowledgments
We would like to thank Helmut Neufeld for useful remarks. D.~M.~was supported by the Austrian 
Science Fund (FWF) under project P17692.

\end{document}